# Subsurface Carbon-Induced Local Charge of Copper for On-Surface Displacement Reaction

Shaoshan Wang[1,4,#], Pengcheng Ding[1,3,#], Zhuo Li[1,4,#], Cristina Mattioli[2], Wenlong E[4], Ye Sun[3], André Gourdon[2], Lev N. Kantorovich[5], Flemming Besenbacher[6], Xueming Yang[4], and Miao Yu[1,4,*]

1. School of Chemistry and Chemical Engineering, Harbin Institute of Technology, Harbin 150001, China

2. CEMES-CNRS, Toulouse 31055, France

3. Condensed Matter Science and Technology Institute, Harbin Institute of Technology, Harbin 150001, China

4. Dalian Institute of Chemical Physics, Chinese Academy of Sciences, Dalian 116023, China

5. Department of Physics, King's College London, The Strand, London WC2R 2LS, United Kingdom

6. iNANO and Department of Physics and Astronomy, Aarhus University, Aarhus 8000, Denmark

# These authors contributed equally to this work.

*Correspondence to: miaoyu_che@hit.edu.cn

**Abstract**

Transition metal carbides have sparked unprecedented enthusiasm as high-performance catalysts in recent years. Still, the catalytic properties of copper (Cu) carbide remain unexplored. By introducing subsurface carbon (C) to Cu(111), displacement reaction of proton in carboxyl acid group with single Cu atom is demonstrated at the atomic scale and room temperature. Its occurrence is attributed to the C-doping induced local charge of surface Cu atoms (up to +0.30 $e$/atom), which accelerates the rate of on-surface deprotonation via reduction of the corresponding energy barrier, thus enabling the instant displacement of a proton with a Cu atom when the molecules land on the surface. Such well-defined and robust $Cu^{\delta+}$ surface based on the subsurface C doping offers a novel catalytic platform for on-surface synthesis.

In recent years, transition metal carbides (TMCs), *e.g.* molybdenum carbide, titanium carbide, *etc.* have shown considerable potential as high-performance catalysts in hydrogen evolution,[1] carbon dioxide reduction,[2] deoxygenation of biomass,[3] and methane dehydroaromatization.[4] TMC catalysts not only accelerate the reaction rate, alter the pathway, but also improve the reaction selectivity,[5] and even enable reactions that cannot be triggered by conventional metal catalysts.[6] However, the reported work has substantially focused on the fabrication of TMCs and their catalytic performance based on spectroscopic measurements, whilst intuitive studies of catalytic reactions on TMCs at the atomic/molecular scales are rather scarce.

Copper (Cu), as one of the most classical catalysts, has been applied extensively in selective oxidation,[7] coupling reaction,[8] carbon dioxide reduction,[9] *etc.* It has been proposed that charge localization on Cu ($Cu^{\delta+}$) achieved by doping nonmetals (*e.g.* oxygen,[9b,10] boron,[11] nitrogen[12]) can significantly boost the catalytic activities of the parent metal and modify the reaction process, *e.g.* enhance the Faradaic efficiency for $C_2$ hydrocarbons in carbon dioxide reduction.[11a] Since the extrinsic species on top of a Cu surface have a high probability to be removed or modified upon chemical reactions, which results in the reduction of $Cu^{\delta+}$ to $Cu^0$,[9b,13] subsurface doping is believed to be advantageous for long-term, stable catalytic performance.[10b,11a] In regard to compatibility with Cu, carbon (C), one of the typical minorities in natural Cu bulk, certainly holds advantages over other elements.[14] However, local charge of Cu surface induced by C-doping and the catalytic properties of Cu carbide or C-doped Cu are still unexplored.

Herein, we report a locally charged Cu surface layer induced by subsurface C (C-$Cu^{\delta+}$) and its catalytic application for an on-surface displacement reaction (Scheme 1). Applying 3,5-*bis*(carboxyl acid)-phenyl-3-maleimide ($C_{12}H_7NO_6$, denoted as BCPM) on C-$Cu^{\delta+}$, efficient substitution of the proton in each carboxylic acid (CA) group of BCPM by a single Cu atom is demonstrated at room temperature (RT), forming a new compound, *i.e.* $Cu_2C_{12}H_5NO_6$. In contrast, such a displacement reaction does not occur on pristine Cu(111), even at elevated temperatures until complete desorption of the adsorbate.[15] The disparity in the two cases is attributed to the local charge of up to +0.30 *e* per Cu atom in the surface layer for C-$Cu^{\delta+}$, which lowers the deprotonation barrier by 0.11 eV (per CA group) hence increases the deprotonation rate significantly. As a result, rapid displacement between protons of CA and Cu atoms is enabled upon the adsorption of BCPM on C-$Cu^{\delta+}$ at RT.

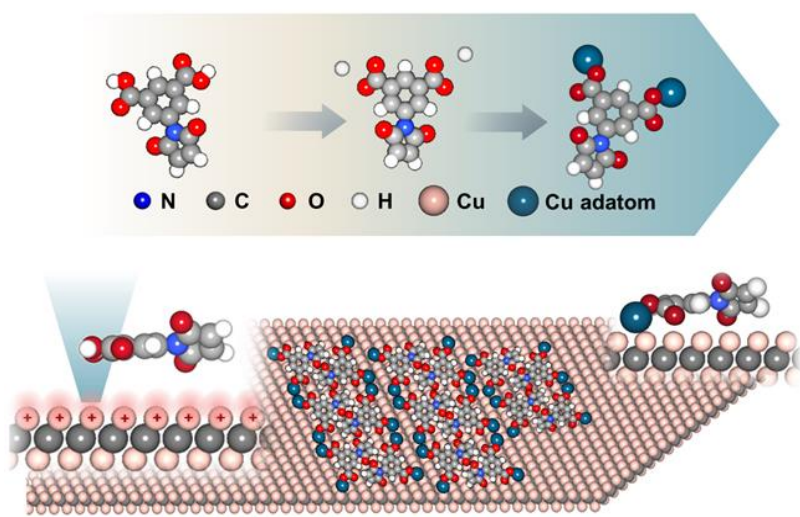

**Scheme 1.** Illustration of positively charged Cu(111) surface induced by subsurface C and its application for on-surface displacement reaction of BCPM compound.

To introduce C to Cu(111), high-temperature decomposition of ethylene was applied [details in the Supporting Information (SI)]. The substrate temperature and ethylene pressure used are much lower than those for graphene production (1020 K *vs.* 1273 K; 1.0 × 10$^{-6}$ mbar *vs.* 1.0 × 10$^{-5}$ mbar).[16] After four cycles of C doping, X-ray photoelectron spectroscopy (XPS) analysis indicates the presence of C. The C *1s$_{1/2}$* XP peak (Figure 1a) contains three components, including a primary one at 283.8 eV (C−Cu bonding) and two minor peaks at 284.5 and 285.3 eV (*sp$^2$*/*sp$^3$* hybridized C).[17] The Cu *2p$_{3/2}$* spectrum (Figure 1b) also reveals the C−Cu bonding (933.1 eV).[18] Importantly, the C atoms are believed to be not located on top of the Cu(111) surface, but diffuse into the substrate, as the Cu terraces show no clusters or flakes of C species in the scanning tunneling microscopy (STM) image (Figure 1c and details in SI), and the binding energy of C−Cu (283.8 eV) in the C *1s$_{1/2}$* XP spectrum is by ~0.6 eV higher than for the on-surface C reported in the literature.[19] The C content can be tuned by varying the number of doping cycles. At higher doses (*e.g.* after eight doping cycles), very likely due to the saturation of subsurface C, small patches of graphene appear on the Cu surface (Figure S1). In this case, the C *1s$_{1/2}$* XP peak at 284.5 eV corresponding to *sp$^2$* hybridized C becomes dominant, the relative intensity of Cu−Cu bonding in the Cu *2p$_{3/2}$* spectrum is much lower (SI, Figure S1).

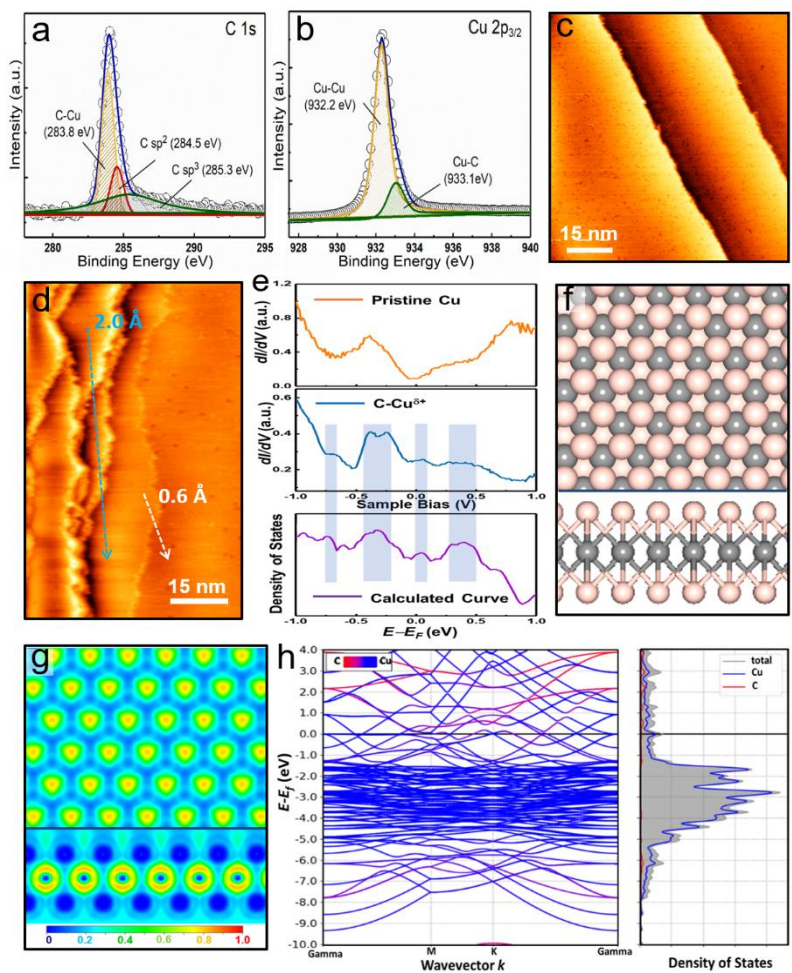

*Figure 1.* a) High-resolution C $1s_{1/2}$ and b) Cu $2p_{3/2}$ XP spectra, and c) large-scale STM image of the Cu substrate after four cycles of C-doping. d) Cu(111) surface treated by two C-doping cycles showing various step heights. e) Normalized dI/dV curve of the pristine and doped Cu(111) (after four doping cycles) together with the calculated LDOS for the model in panel f. Top and side views of f) structural model and g) ELF mapping, and h) calculated band structure and DOS of the C-doped Cu(111) with all subsurface octahedron sites occupied by C atoms. Only top two Cu layers are actually shown for clarity. In panel e, Cu and C atoms are in pink and grey, respectively.

Compared with pristine Cu(111), C-doped Cu surface shows barely modified atomic arrangement with smooth step edges as well as small, shallow pits randomly distributed on the surface (Figure 1c and S2). Interestingly, at low dosages, low steps of ~0.6 Å and high steps of ~2.6 Å are observed throughout the image (Figures 1d and S3). Bear in mind that the step height of pristine Cu(111) is 2.0 Å. Moreover, the *dI/dV* curve after C doping has changed. As depicted in Figure 1e (collected from the marked position in Figure S2d), pristine Cu(111) shows an evident peak at −0.41 V, whereas the C-doped surface presents the main peak at −0.32 V and minor peaks at −0.73, 0.04 and 0.42 V.

Consistent with our STM/XPS and earlier theoretical results,[20] DFT calculations reveal that the most stable occupation for the C atoms is the subsurface octahedron site. We optimized four substrate slabs: each involving five Cu layers and one C layer just below the outermost Cu layer, with one (1C), two (2C), three (3C) or four C atoms (4C) in a 2×2 cell (Figure 1f and S4). The 4C model corresponds to the case when all subsurface octahedron sites are occupied by C. The total formation energy and the accumulated charge on C and its neighboring Cu atoms are varied with the C content (Table S1). In the 4C model (Figure S5), distinct structural relaxation occurs: the interlayer spacing of the top four Cu layers is modified due to the presence of the C layer while the lateral Cu−Cu distance remains unchanged. The top two Cu layers spacing is enlarged by 0.81 Å (see SI). This model is in good agreement with the STM results, which show unchanged lateral atomic arrangement and varied step height (Figure S2 and S3). The calculated local density of states (LDOS) based on the 4C model matches the dI/dV curve of the C-doped surface (Figures 1e and S6). All these STM, STS, XPS and DFT results thus indicate the formation of subsurface C. The penetration of C into Cu substrate is determined by the thermodynamics of the doping process (Figure S7, detailed discussion in SI).

In the 4C model, C accepts electrons from the surrounding Cu (Figure S8), resulting in a local charge of +0.30 $e$ and −0.62 $e$ on each surface Cu and each subsurface C. The positively/ negatively charged Cu/C is further confirmed by the intuitive mapping based on electron localization function (ELF, Figure 1g), indicating the significant C−Cu bonding.[21] The Cu(111) surface with the saturated C subsurface layer is then denoted as '4*C-Cu$^{\delta+}$*'. The calculated electronic structure indicates that this C-doped surface is metallic (Figure 1h, detailed discussion in SI).

Superiority of the *C-Cu$^{\delta+}$* surface was then demonstrated by on-surface reaction of BCPM compound. We first investigated the assembly of BCPM on pristine Cu(111). As shown in our earlier work,[15] BCPM molecules form ordered domains composed of parallel '*zippers*' along the <1-10> direction of Cu(111) at RT (Figure S9), denoted as '*Z1*' structure. Each single BCPM is imaged as an inverted Erlenmeyer flask decorated by a spherical bell, corresponding to the *bis*(carboxyl acid)-phenyl and maleimide groups, respectively. Each zipper is composed of two interlocked molecular rows, with a periodicity along the rows of 10.2 ± 0.2 Å. The axis-to-axis distance between the zippers is 15.8 ± 0.3 Å. The CA group is believed to be intact (*i.e.* without deprotonation) in this case, given that (1) the calculated STM image for the monolayer of intact BCPM fits the experimental observations better than the images for the deprotonated case (Figure S10), and (2) the O 1$s_{1/2}$ XP results below suggests so.

When BCPM molecules are deposited on the C-doped Cu(111) at RT, a different well-defined structure is formed (Figure 2a), denoted as '*Z2*'. Similar to *Z1*, each domain is also composed of parallel zippers of interlocked molecular double rows along <1-10> direction of Cu(111), with the periodicity along the row of 10.2 ± 0.2 Å. However, the inter-*zipper* periodicity increases to 17.5 ± 0.3 Å, with additional oval protrusions (OPs) (marked by the white ovals in Figure 2b–c) also clearly visible. These OPs are located between CA groups of two adjacent molecules, and decorate both sides of each *zipper*. At a relatively low C content, *Z2* is observed to coexist with *Z1*, even within a single molecular domain (Figure S11); by increasing the C content, the *Z2* structure can coexist with small graphene patches, with no *Z1*. Purposely roughened pristine Cu(111) with high density of surface defects cannot induce *Z2* either (SI, Figure S12).

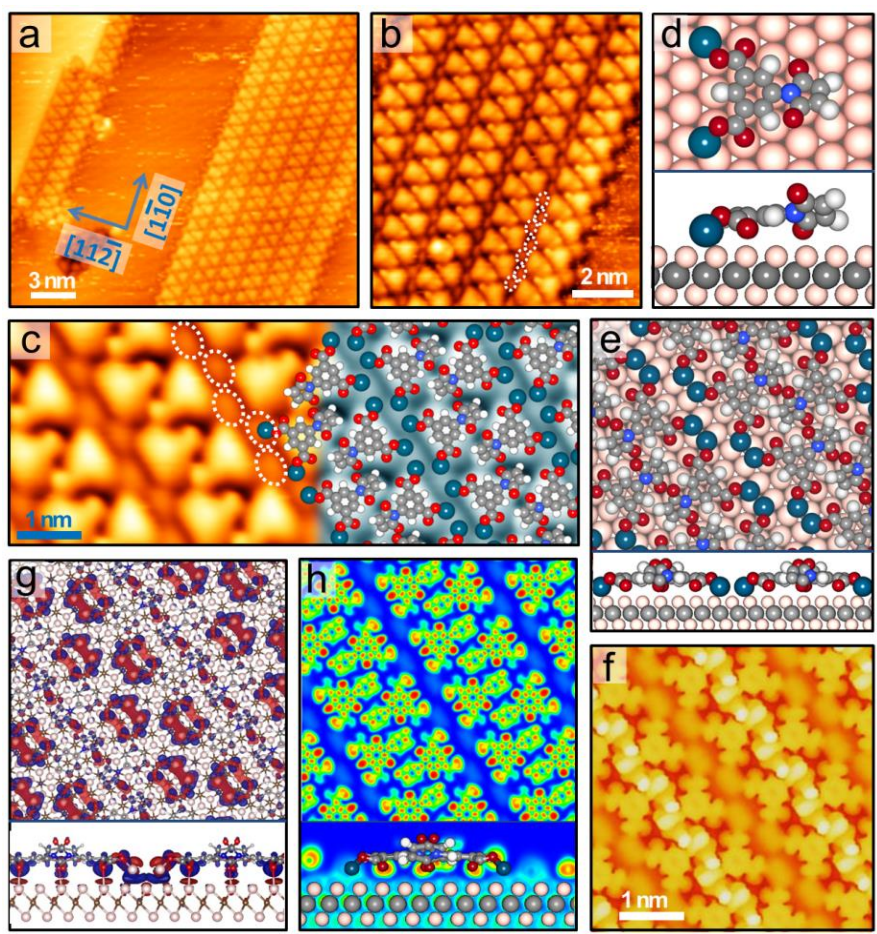

*Figure 2.* a) Large-scale and b-c) close-view STM images of the well-ordered '*Z2*' structure, with additional protrusions (marked by the white dashed ovals) attached on both sides of the molecular *zippers*. d) Top and side views of a single *DP-2Cu* on the 4*C*-$Cu^{\delta+}$ surface modeled by DFT, where each COO− group is rotated by 27° and the maleimide group is rotated by 37° relative to the surface. e) DFT optimized structure of *Z2* on 4*C*-$Cu^{\delta+}$ and f) simulated STM image based on this model, showing a good match with the experimental observations. g) Top and side views of charge density difference plot (the red and blue colors indicate charge depletion and accumulation, the isosurface value is 0.02 e Å$^{-3}$) and h) ELF mapping of *DP-2Cu zippers* on the 4*C*-$Cu^{\delta+}$ surface, indicating the charge transfer from the Cu substrate to $O_{CO}$ and $O_{ML}$.

We assign the OPs to Cu atoms, given that (1) the OPs have similar tunneling contrast to the Cu substrate, which is much lower than that of the molecules; (2) no impurities or molecular fractures are observed under the same BCPM deposition conditions on pristine Cu(111); (3) the abundance of free Cu atoms on Cu(111) is well known.[22] Four structural models were considered for *Z2* with 4C-Cu$^{\delta+}$ as the substrate: (1) intact BCPM with each OP assigned to one Cu atom ('*BCPM-1Cu*') (Figure S13a); (2) BCPM with deprotonated (DP) CA and each OP assigned to one Cu atom ('*DP-1Cu*') (Figure S13b); (3) intact BCPM with each OP assigned to two Cu atoms ('*BCPM-2Cu*') (Figure S13c); (4) BCPM with deprotonated CA and each OP assigned to two Cu atoms ('*DP-2Cu*') (Figure 2d–e). Clearly, only the model of *DP-2Cu*, *i.e.* BCPM molecule with the two protons in CA displaced by two Cu atoms, matches the experimental observations (Figure 2f).

As shown in Figure 2d, for each single *DP-2Cu* molecule, Cu–O bond length is 1.9 Å, and the two Cu atoms in the two COO–Cu groups are located on a three-fold hollow site at the height of 2.0 Å above the surface. While the phenyl ring remains flat, the two COO– groups are rotated by 27° so that the O in C=O ('$O_{CO}$') sits exactly atop of a Cu substrate atom at a distance of 2.2 Å. The maleimide group is rotated by 37° relative to the Cu surface with an $O_{ML}$–Cu substrate distance of 2.1 Å. The *zippers* composed of *DP-2Cu* are packed side-by-side along the <1-10> direction of Cu(111) (Figure 2e), with the inner-row and inter-zipper periodicities in a good accordance with the observed '*Z2*' structure. The molecular morphology, contrast and arrangement in the simulated STM image (Figure 2f) based on this structural model are all consistent with the experimental observations. Revealed by the charge density difference plot (Figure 2g) and ELF mapping (Figure 2h), $O_{CO}$ and $O_{ML}$ accept 0.13 *e* and 0.10 *e* from the Cu substrate atoms, whilst the rest atoms in *DP-2Cu* have no charge transfer with the substrate.

The displacement reaction of BCPM was further ascertained by XPS results. O $1s_{1/2}$ XP spectrum of BCPM on pristine Cu(111) shows three primary sub-peaks located at 532.0, 532.1 and 533.3 eV (Figure 3a), corresponding to carbonyl O in intact CA,[23] carbonyl O in maleimide ring[24] and hydroxyl O in intact CA,[23a–b,25] respectively, with the area ratio of 0.9:1:1. In addition, there is a tiny component located at 530.9 eV, which may be attributed to the BCPMs with deprotonated CA. According to the earlier reports,[26] deprotonation of CA group could occur at the step edges of Cu(111) at room temperature. Given its very low intensity, the result indicates that most of BCPMs remain intact on pristine Cu(111) at RT. The spectrum of BCPM on the C-doped Cu includes three components located at 530.9, 531.2, and 532.1 eV with an identical integral area. The peak at 530.9 eV is attributed to oxygen in carboxylate[23c,27] (COO–Cu here). The peak at 531.2 eV is assigned to carbonyl O in the reacted CA. Due to the bonding with Cu, the reacted CA rotates so that its carbonnyl O is close to the substrate and receives an electronic denisty from Cu (Figures 2d and 2g), resulting in the shift to the lower binding energy.The results support our interpretation of the substitution of the proton in CA by Cu for *Z2*, while the molecules remain intact in *Z1*.

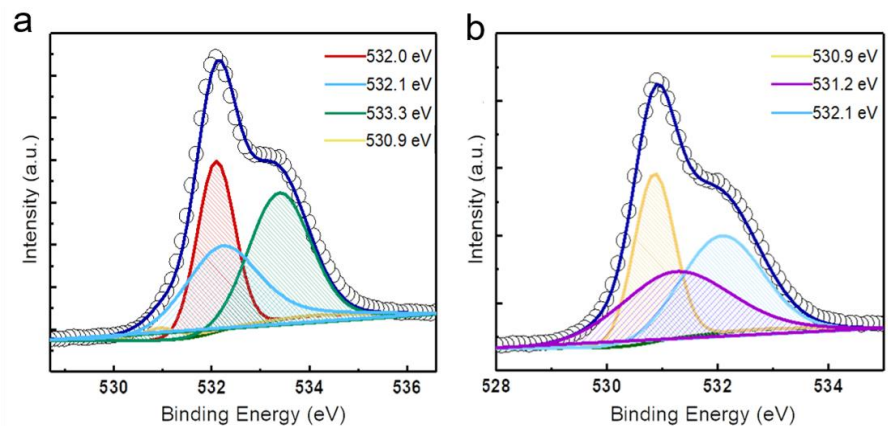

*Figure 3.* High-resolution O $1s_{1/2}$ XP spectra of BCPM on a) pristine Cu(111) and b) Cu(111) treated with four C-doping cycles.

Very interestingly, although the deprotonation of the CA groups is expected to occur on pristine Cu(111) at elevated temperature according to the literature,[26] *Z2* structure is not observed (without C-doping) up to the temperature at which the adsorbates completely desorb from the surface.[15] The influence of the positive local charge of Cu induced by the subsurface C on the displacement reaction was then explored further, by comparing the deprotonation process for a model system of a single benzyl acid molecule on pristine Cu(111) with that on the *4C-Cu$^{\delta+}$* surface (details in SI).

Distinct from the case on the pristine surface, the positive charge localized on the surface Cu for *4C-Cu$^{\delta+}$* attracts the negatively charged O in CA, whereby facilitating the breakage of O−H and formation of the Cu−O bonding. It is found that the deprotonation barrier on *4C-Cu$^{\delta+}$* decreases by 0.11 eV compared with that on pristine Cu(111). Whilst the reaction on the pristine Cu is endothermic with its deprotonated final state (FS) being 0.27 eV higher than its initial state (IS, Figure S14a–b), the FS on *4C-Cu$^{\delta+}$* is by 0.24 eV lower than its IS (Figure S14b–c). Consequently, the calculated rate constant for the deprotonation of the CA group increases significantly on *4C-Cu$^{\delta+}$* at 300 K (SI), as compared with the pristine surface. As a result, the displacement reaction becomes highly favored on the *4C-Cu$^{\delta+}$* surface upon BCPM deposition. At variance, on the pristine Cu(111), BCPM only forms *Z1* structure upon adsorption, where the close-packed arrangement and the significant ionic molecule-substrate interaction impede the diffusion and attachment of Cu adatoms even when the CA would deprotonate at a higher temperature.

In summary, by introducing subsurface C to Cu(111), efficient displacement reaction of proton in CA of BCPM and Cu atom is enabled, which does not occur on the pristine substrate. This is attributed to the positive charge localized on surface Cu atoms when introducing subsurface C, which lowers its barrier, makes the reaction exothermic, and increases the deprotonation rate. The displacement reaction is thus highly favorable upon adsorption of BCPM on the *C-Cu$^{\delta+}$* surface at RT. This work not only reports for the first time the charge localization and the catalytic application of the C-doped Cu, but also provides a novel platform for the on-surface synthesis. Superior to the commonly-used high surface-energy metal crystal planes or doped metal surfaces (that are significantly roughened with various islands/holes), the well-defined *C-Cu$^{\delta+}$* surface offers a much less anisotropic platform for long-range ordered synthesis. Moreover, given the high-temperature and subsurface doping, the platform is robust upon thermal excitation and avoids the reduction to Cu$^0$ by the on-surface reactions. Such locally charged transition metal surfaces therefore hold great promise for on-surface synthesis and high-performance catalysis.

**References**

[1]  a) L. Lin, W. Zhou, R. Gao, S. Yao, X. Zhang, W. Xu, S. Zheng, Z. Jiang, Q. Yu, Y. W. Li, C. Shi, X. D. Wen, D. Ma, *Nature* **2017**, *544*, 80–83; b) S. Yao, X. Zhang, W. Zhou, R. Gao, W. Xu, Y. Ye, L. Lin, X. Wen, P. Liu, B. Chen, E. Crumlin, J. Guo, Z. Zuo, W. Li, J. Xie, L. Lu, C. J. Kiely, L. Gu, C. Shi, J. A. Rodrigue, D. Ma, *Science* **2017**, *357*, 389–393; c) Q. Gao, W. Zhang, Z. Shi, L. Yang, Y. Tang, *Adv. Mater.* **2019**, *31*, 1802880; d) Y. T. Xu, X. Xiao, Z. M. Ye, S. Zhao, R. Shen, C. T. He, J. P. Zhang, Y. Li, X. M. Chen, *J. Am. Chem. Soc.* **2017**, *139*, 5285–5288.

[2]  a) M. D. Porosoff, X. Yang, J. A. Boscoboinik, J. G. Chen, *Angew. Chem. Int. Ed.* **2014**, *53*, 6705–6709; b) J. Wang, S. Kattel, C. J. Hawxhurst, J. H. Lee, B. M. Tackett, K. Chang, N. Rui, C. J. Liu, J. G. Chen, *Angew. Chem. Int. Ed.* **2019**, *58*, 6271–6275; c) Y. L. Men, Y. You, Y. X. Pan, H. Gao, Y. Xia, D. G. Cheng, J. Song, D. X. Cui, N. Wu, Y. Li, S. Xin, J. B. Goodenough, *J. Am. Chem. Soc.* **2018**, *140*, 13071–13077; d) S. Yao, B. Yan, Z. Jiang, Z. Liu, Q. Wu, J. H. Lee, J. G. Chen, *ACS Catal.* **2018**, *8*, 5374–5381.

[3]  a) W. Wan, S. C. Ammal, Z. Lin, K. E. You, A. Heyden, J. G. Chen, *Nat. Commun.* **2018**, *9*, 4612; b) K. Murugappan, E. M. Anderson, D. Teschner, T. E. Jones, K. Skorupska, Y. Román-Leshkov, *Nat. Catal.* **2018**, *1*, 960–967; c) R. W. Gosselink, D. R. Stellwagen, J. H. Bitter, *Angew. Chem. Int. Ed.* **2013**, *52*, 5089–5092.

[4]  a) H. F. Li, L. X. Jiang, Y. X. Zhao, Q. Y. Liu, T. Zhang, S. G. He, *Angew. Chem. Int. Ed.* **2018**, *130*, 2662–2666; b) C. Geng, T. Weiske, J. Li, S. Shaik, H. Schwarz, *J. Am. Chem. Soc.* **2019**, *141*, 599–610.

[5]  a) J. G. Chen, *Chem. Rev.* **1996**, *96*, 1477–1498; b) W. Wan, B. M. Tackett, J. G. Chen, *Chem. Soc. Rev.* **2017**, *46*, 1807–1823; c) D. W. Flaherty, S. P. Berglund, C. B. Mullins, *J. Catal.* **2010**, *269*, 33–43.

[6] a) T. S. Khan, S. Balyan, S. Mishra, K. K. Pant, M. A. Haider, *J. Phys. Chem. C* **2018**, *122*, 11754–11764; b) I. Lezcano-González, R. Oord, M. Rovezzi, P. Glatzel, S. W. Botchway, B. M. Weckhuysen, A. M. Beale, *Angew. Chem. Int. Ed.* **2016**, *55*, 5215–5219; c) J. Gao, Y. Zheng, J. M. Jehng, Y. Tang, I. E. Wachs, S. G. Podkolzin, *Science* **2015**, *348*, 686–690.

[7] a) J. Baek, B. Rungtaweevoranit, X. Pei, M. Park, S. C. Fakra, Y. S. Liu, R. Matheu, S. A. Alshmimri, S. Alshihri, C. A. Trickett, G. A. Somorjai, O. M. Yaghi, *J. Am. Chem. Soc.* **2018**, 140, 18208–18216; b) V. L. Sushkevich, D. Palagin, M. Ranocchiari, J. A. van Bokhoven, *Science* **2017**, *356*, 523–527.

[8] a) M. Lepper, J. Koebl, L. Zhang, M. Meusel, H. Hoelzel, D. Lungerich, N. Jux, A. de Siervo, B. Meyer, H. P. Steinrueck, H. Marbach, *Angew. Chem. Int. Edit.* **2018**, *57*, 10074–10079; b) P. Leophairatana, S. Samanta, C. C. De Silva, J. T. Koberstein, *J. Am. Chem. Soc.* **2017**, *139*, 3756–3766.

[9] a) A. Loiudice, P. Lobaccaro, E. A. Kamali, T. Thao, B. H. Huang, J. W. Ager, R. Buonsanti, *Angew. Chem. Int. Ed.* **2016**, *55*, 5789–5792; b) H. Mistry, A. S. Varela, C. S. Bonifacio, I. Zegkinoglou, I. Sinev, Y. W. Choi, K. Kisslinger, E. A. Stach, J. C. Yang, P. Strasser, B. Roldan, *Nat. Commun.* **2016**, *7*, 12123; c) C. T. Dinh, T. Burdyny, M. G. Kibria, A. Seifitokaldani, C. M. Gabardo, F. P. G. de Arquer, A. Kiani, J. P. Edwards, P. De Luna, O. S. Bushuyev, C. Zou, R. Quintero-Bermudez, D. Sinton, E. H. Sargent, *Science* **2018**, *360*, 783–787.

[10] a) F. S. Roberts, K. P. Kuhl, A. Nilsson, *Angew. Chem. Int. Edit.* **2015**, *54*, 5179–5182; b) M. Favaro, H. Xiao, T. Cheng, W. A. Goddard, III, J. Yano, E. J. Crumlin, *Proc. Natl. Acad. Sci. USA.* **2017**, *114*, 6706–6711; c) H. Xiao, W. A. Goddard, T. Cheng, Y. Liu, *Proc. Natl. Acad. Sci. USA.* **2017**, *114*, 6685–6688.

[11] a) Y. Zhou, F. Che, M. Liu, C. Zou, Z. Liang, P. De Luna, H. Yuan, J. Li, Z. Wang, H. Xie, H. Li, P. Chen, E. Bladt, R. Quintero-Bermudez, T. K. Sham, S. Bals, J. Hofkens, D. Sinton, G. Chen, E. H. Sargent, *Nat. Chem.* **2018**, *10*, 974–980; b) C. Chen, X. Sun, L. Lu, D. Yang, J. Ma, Q. Zhu, Q. Qian, B. Han, *Green Chem.* **2018**, *20*, 4579–4583.

[12] Z. Q. Liang, T. T. Zhuang, A. Seifitokaldani, J. Li, C. W. Huang, C. S. Tan, Y. Li, P. De Luna, C. T. Dinh, Y. Hu, Q. Xiao, P. Hsieh, Y. Wang, F. Li, R. Quintero-Bermudez, Y. Zhou, P. Chen, Y. Pang, S. C. Lo, L. J. Chen, H. Tan, Z. Xu, S. Zhao, D. Sinton, E. H. Sargent, *Nat. Commun.* **2018**, *9*. 3828.

[13] a) S. Lee, D. Kim, J. Lee, *Angew. Chem. Int. Ed.* **2015**, *54*, 14701–14705; b) D. Ren, Y. Deng, A. D. Handoko, C. S. Chen, S. Malkhandi, B. S. Yeo, *ACS Catal.* **2015**, *5*, 2814–2821; c) A. Dutta, M. Rahaman, N. C. Luedi, P. Broekmann, *ACS Catal.* **2016**, *6*, 3804–3814; d) Y. Lum, J. W. Ager, *Angew. Chem. Int. Ed.* **2018**, *57*, 551–554.

[14] J. Xiao, A. Kuc, T. Frauenheim, T. Heine, *J. Mater. Chem. A* **2014**, *2*, 4885–4889.

[15] S. Wang, Z. Li, P. Ding, C. Mattioli, W. Huang, Y. Wang, A. Gourdon, Y. Sun, M. Chen, L. Kantorovich, X. Yang, F. Rosei, M. Yu, *Angew. Chem. Int. Ed.* **2021**, DOI: 10.1002/anie.202106477.

[16] L. Gao, J. R. Guest, N. P. Guisinger, *Nano Lett.* **2010**, *10*, 3512–3516.

[17] a) Y. Li, Z. Li, A. Ahsen, L. Lammich, G. J. Mannie, J. H. Niemantsverdriet, J. V. Lauritsen, *ACS Catal.* **2018**, *9*, 1264–1273; b) S. Hofmann, R. Sharma, C. Ducati, G. Du, C. Mattevi, C. Cepek, M. Cantoro, S. Pisana, A. Parvez, F. Cervantes-Sodi, A. C. Ferrari, R. Dunin-Borkowski, S. Lizzit, L. Petaccia, A. Goldoni, J. Robertson, *Nano Lett.* **2007**, *7*, 602–608; c) R. S. Weatherup, H. Amara, R.

Blume, B. Dlubak, B. C. Bayer, M. Diarra, M. Bahri, A. Cabrero-Vilatela, S. Caneva, P. R. Kidambi, M. B. Martin, C. Deranlot, P. Seneor, R. Schloegl, F. Ducastelle, C. Bichara, *J. Am. Chem. Soc.* **2014**, *136*, 13698–13708.

[18] B. Balamurugan, B. R. Mehta, S. M. Shivaprasad, *Appl. Phys. Lett.* **2002**, *82*, 115–117.

[19] Q. Sun, L. Cai, S. Wang, R. Widmer, H. Ju, J. Zhu, L. Li, Y. He, P. Ruffieux, R. Fasel, W. Xu, *J. Am. Chem. Soc.* **2016**, *138*, 1106–1109.

[20] a) S. Riikonen, A. V. Krasheninnikov, L. Halonen, R. M. Nieminen, *J. Phys. Chem. C* **2012**, *116*, 5802–5809; b) O. V. Yazyev, A. Pasquarello, *Phys. Rev. Lett.* **2008**, *100*. 156102.

[21] a) Q. Sun, Q. Wang, J. Z. Yu, V. Kumar, Y. Kawazoe, *Phys. Rev. B* **2001**, *63, 193408*; b) S. Zhang, Q. Wang, Y. Kawazoe, P. Jena, *J. Am. Chem. Soc.* **2013**, *135*, 18216–18221.

[22] a) N. Lin, D. Payer, A. Dmitriev, T. Strunskus, C. Woll, J. V. Barth, K. Kern, *Angew. Chem. Int. Edit.* **2005**, *44*, 1488–1491; b) F. Bebensee, K. Svane, C. Bombis, F. Masini, S. Klyatskaya, F. Besenbacher, M. Ruben, B. Hammer, T. R. Linderoth, *Angew. Chem. Int. Edit.* **2014**, *53*, 12955–12959; c) Q. Li, B. Yang, J. Bjork, Q. Zhong, H. Ju, J. Zhang, N. Cao, Z. Shi, H. Zhang, D. Ebeling, A. Schirmeisen, J. Zhu, L. Chi, *J. Am. Chem. Soc.* **2018**, 140, 6076–6082.

[23] a) L. Wu, Z. Cai, M. Liu, W. Ye, H. Ju, J. Zhu, D. Zhong, *J. Phys. Chem. C* **2018**, *122*, 9075—9080; b) H.Y. Gao, M. Sekutor, L. Liu, A. Timmer, H. Schreyer, H. Moenig, S. Amirjalayer, N. A. Fokina, A. Studer, P. R. Schreiner, H. Fuchs, *J. Am. Chem. Soc.* **2019**, *141*, 315–322; c) C. Morchutt, J. Björk, C. Straßer, U. Starke, R. Gutzler, K. Kern, *ACS Nano* **2016**, *10*, 11511—11518.

[24] a) G. H. De Zoysa, V. Sarojini, *ACS Appl. Mater. Interfaces,* **2017***, 9,* 1373–1383; b) C. S. Park, H. J. Lee, A. C. Jamison, T. R. Lee, *Langmuir* **2016**, *32*, 7306–7315.

[25] J. I. Urgel, B. Cirera, Y. Wang, W. Auwaerter, R. Otero, J. M. Gallego, M. Alcami, S. Klyatskaya, M. Ruben, F. Martin, R. Miranda, D. Ecija, J. V. Barth, *Small* **2015**, *11*, 6358—6364.

[26] T. Schmitt; L. Hammer; M. A. Schneider, *J. Phys. Chem. C* **2016**, *120*, 1043–1048.

[27] a) T. Classen, M. Lingenfelder, Y. Wang, R. Chopra, C. Virojanadara, U. Starke, G. Costantini, G. Fratesi, S. Fabris, S. de Gironcoli, S. Baroni, S. Haq, R. Raval, K. Kern, *J. Phys. Chem. A* **2007**, *111*, 12589–12603; b) Y. Wang, S. Fabris, T. W. White, F. Pagliuca, P. Moras, M. Papagno, D. Topwal, P. Sheverdyaeva, C. Carbone, M. Lingenfelder, T. Classen, K. Kern, G. Costantini, *Chem. Commun.* **2012**, *48*, 534–536.

**Acknowledgements:** This work is financially supported by the National Natural Science Foundation of China (21473045, 51772066), and State Key Laboratory of Urban Water Resource and Environment, Harbin Institute of Technology (2021TS08).

**Conflict of interest:** The authors declare no conflict of interest.